\documentclass[conference,final]{IEEEtran}
\IEEEoverridecommandlockouts

\usepackage{amsmath,amssymb,amsfonts}
\usepackage{algorithmic}
\usepackage{graphicx}
\usepackage{textcomp}

\usepackage[
  unicode=true,
  pdfencoding=unicode,
  plainpages=false,
  pdfpagelabels,
  breaklinks=true,
  pdfauthor={Renato Cordeiro Ferreira},
  pdftitle={A Metrics-Oriented Architectural Model to Characterize Complexity on Machine Learning-Enabled Systems},
  pdfkeywords={Software Metrics, Software Complexity, ML-Enabled Systems, Machine Learning Engineering, MLOps},
]{hyperref}

\usepackage[svgnames,x11names,table]{xcolor}

\hypersetup{
  colorlinks=true,
  citecolor=DarkGreen,
  linkcolor=NavyBlue,
  urlcolor=DarkRed,
  filecolor=green,
  anchorcolor=black,
}

\usepackage[nameinlink,noabbrev]{cleveref}

\usepackage{mdframed}

\usepackage{enumitem}

\usepackage{tikz}
\usetikzlibrary{arrows.meta}

\usepackage[style=ieee,url=false,isbn=false,doi=false,eprint=false]{biblatex}
\newcommand{\Quote}[1]{\textit{``#1''}}

\mdfdefinestyle{ResearchQuestion}{%
    linecolor=black,
    outerlinewidth=8pt,
    skipabove=\baselineskip,
    skipbelow=\baselineskip,
    innertopmargin=8pt,
    innerbottommargin=8pt,
    innerrightmargin=8pt,
    innerleftmargin=8pt,
}

\newcounter{RQ}
\renewcommand{\theRQ}{\arabic{RQ}}

\newcounter{SubRQ}[RQ] 
\renewcommand{\theSubRQ}{\theRQ.\arabic{SubRQ}}

\newenvironment{researchquestion}{%
  \refstepcounter{RQ}%
  \begin{mdframed}[style=ResearchQuestion]\label{rq:\theRQ}%
  \textbf{RQ\theRQ:}%
}{%
  \end{mdframed}%
}

\mdfdefinestyle{RevisitResearchQuestion}{%
    linecolor=black,
    outerlinewidth=8pt,
    skipabove=\parskip,
    skipbelow=\parskip,
    innertopmargin=8pt,
    innerbottommargin=8pt,
    innerrightmargin=8pt,
    innerleftmargin=8pt,
}

\crefformat{RQ}{#2RQ#1#3}
\Crefformat{RQ}{#2RQ#1#3}
\crefformat{SubRQ}{#2RQ#1#3}
\Crefformat{SubRQ}{#2RQ#1#3}

\mdfdefinestyle{ExpectedResult}{%
    linecolor=black,
    outerlinewidth=8pt,
    skipabove=\baselineskip,
    skipbelow=\baselineskip,
    innertopmargin=8pt,
    innerbottommargin=8pt,
    innerrightmargin=8pt,
    innerleftmargin=8pt,
}

\newcounter{ER}
\renewcommand{\theER}{\arabic{ER}}

\newenvironment{expectedresult}{%
  \refstepcounter{ER}%
  \begin{mdframed}[style=ExpectedResult]\label{er:\theER}%
  \textbf{ER\theER:}%
}{%
  \end{mdframed}%
}

\crefformat{ER}{#2ER#1#3}
\Crefformat{ER}{#2ER#1#3}

\newcounter{MethPhase}
\renewcommand{\theMethPhase}{\arabic{MethPhase}}

\newcounter{MethStep}[MethPhase] 
\renewcommand{\theMethStep}{\theMethPhase.\arabic{MethStep}}

\newcommand{\MethodologyPhase}{\refstepcounter{MethPhase}\label{phase:\theMethPhase}}

\newcommand{\MethodologyStep}{\refstepcounter{MethStep}\label{step:\theMethStep}}

\crefname{MethPhase}{phase}{phases}
\Crefname{MethPhase}{Phase}{Phases}
\crefname{MethStep}{step}{steps}
\Crefname{MethStep}{Step}{Steps}

\definecolor{drawio-blue}{HTML}{1ba1e2} %
\definecolor{drawio-gray}{HTML}{647687} %
\definecolor{drawio-green}{HTML}{6d8764} %
\definecolor{drawio-orange}{HTML}{fa6800} %
\definecolor{drawio-pink}{HTML}{99004D} %
\definecolor{drawio-purple}{HTML}{76608a} %
\definecolor{drawio-red}{HTML}{e51400} %
\definecolor{drawio-white}{HTML}{f9f7ed} %
\definecolor{drawio-black}{HTML}{36393d} %
\definecolor{drawio-yellow}{HTML}{e3c800} %

\definecolor{drawio-violet}{HTML}{6a00ff} %
\definecolor{drawio-magenta}{HTML}{dd0073} %
\definecolor{drawio-moss}{HTML}{008a00} %

\newcommand{\TikzSkewedSquare}[1][black]{%
  \tikz[baseline] {\draw[transform shape, black, fill={#1}] (0,0) rectangle (1.5ex,1.5ex);}%
}
\newcommand{\TikzSkewedCircle}[1][black]{%
  \tikz[baseline] {\draw[yshift=0.75ex, anchor=south, transform shape, black, fill={#1}] (0,0) circle (0.75ex);}%
}

\newcommand{\LegendColoredSquare}[3]{%
  \textcolor{#1}{#3}~\TikzSkewedSquare[#2]%
}
\newcommand{\LegendColoredCircle}[3]{%
  \textcolor{#1}{#3}~\TikzSkewedCircle[#2]%
}

\newcommand{\LegendColoredComponent}[2]{%
  \LegendColoredSquare{drawio-#1}{drawio-#1}{#2}%
}
\newcommand{\LegendBWComponent}[1]{%
  \LegendColoredSquare{drawio-black}{drawio-white}{#1}%
}
\newcommand{\LegendColoredLabel}[2]{%
  \LegendColoredCircle{drawio-#1}{drawio-#1}{#2}%
}

\begin{document}

\title{
A Metrics-Oriented Architectural Model\break
to Characterize Complexity on\break
Machine Learning-Enabled Systems
\thanks{This study was financed in part by the Coordenação de Aperfeiçoamento
de Pessoal de Nível Superior -- Brasil (CAPES) -- Finance Code 001. It also
received support from the MARIT-D project, co-funded from the Internal
Security Fund -- Police programme under grant agreement no. 101114216.}
}

\author{%
\IEEEauthorblockN{Renato Cordeiro Ferreira}
\IEEEauthorblockA{\textit{Institute of Mathematics and Statistics} \\
\textit{University of São Paulo}\\
São Paulo, Brazil \\
0000-0001-7296-7091}
}

\maketitle

\begin{abstract}
How can the complexity of ML-enabled systems be managed effectively?
The goal of this research is to investigate how complexity affects
ML-Enabled Systems (MLES). To address this question, this research
aims to introduce a metrics-based architectural model to characterize
the complexity of MLES. The goal is to support architectural decisions,
providing a guideline for the inception and growth of these systems.
This paper showcases the first step for creating the metrics-based
architectural model: an extension of a reference architecture that
can describe MLES to collect their metrics.
\end{abstract}

\begin{IEEEkeywords}
Software Metrics, Software Complexity, ML-Enabled Systems,
Machine Learning Engineering, MLOps.
\end{IEEEkeywords}

\section{Introduction}\label{sec:introduction}

  \emph{Complexity} has been a subject of discussion since the early
  days of the Software Engineering field~\parencite{Brooks1975TheMan-Month}.
  In the book~\citetitle{Brooks1975TheMan-Month},
  \citeauthor*{Brooks1975TheMan-Month} introduces the concept of
  \emph{essential} and \emph{accidental} complexity for software%
  ~\parencite{Brooks1975TheMan-Month}:
    the \emph{essential} exists intrinsically to satisfy the requirements
    of the problem solved, whereas the \emph{accidental} may originate
    from any external factors that influence the solution chosen.
  Under this definition, an ML-enabled system (MLES) has high essential
  complexity: it requires extra components in its architecture to
  support data processing and model handling%
  ~\parencite{Ameisen2020BuildingApplications, Amershi2019SoftwareStudy,
  Benton2020MachineApplications}.
  
  According to Gartner's report~\parencite{Gartner2022},
  only around 54\% of AI projects successfully reach production.
  Besides technical challenges~\parencite{Sculley2015HiddenSystems},
  many factors influence the ability to deliver, such as
    infrastructure,
    developer experience,
    development processes, and
    team composition.
  All of them reflect into the software architecture%
  ~\parencite{Brooks1975TheMan-Month}, thus becoming potential
  sources of accidental complexity.
  
  
  This PhD research outlines \emph{a metrics-oriented architectural
  model to characterize the complexity of ML-enabled systems}.
  The goal is to use metrics to identify where complexity
  emerges in the software architecture of MLES.
  Hopefully, this technique will assist developers to manage
  complexity, allowing them to reach production more often.
  
  \section{Related Literature}

  While there are many recent surveys summarizing important gaps in the SE4AI
  literature, only \citeauthor*{Giray2021AChallenges} highlights \emph{handling
  complexity} as an open challenge for MLE~\parencite{Giray2021AChallenges}.
  However, this idea is also supported by
  \citeauthor*{Tamburri2020SustainableChallenges}, who outlines challenges for
  sustainable AI~\parencite{Tamburri2020SustainableChallenges}.
  \mbox{In an interview} study with 18 professional ML engineers%
  ~\parencite{Shankar2022OperationalizingStudy},
  \citeauthor*{Shankar2022OperationalizingStudy} paper describes how
  participants \Quote{expressed an aversion to complexity}, implying
  that practitioners consider it an issue.
  
  The survey by \citeauthor*{Diaz-De-Arcaya2023ASurvey} further
  explores possible origins of complexity in MLES%
  ~\parencite{Diaz-De-Arcaya2023ASurvey}, which become visible in
  the software architecture:
  \begin{itemize}
  \item they are harder to modularize,
        since data modeling and data processing are highly coupled%
        ~\parencite{Wan2021HowPractices};
  \item they require an understanding of ML principles and techniques, 
        with implications to the number of ML models required, their size,
        versioning, and tracking~\parencite{Lwakatare2020Large-ScaleSolutions,Priestley2023APipelines};
  \item they often handle big data,
        which leads to the adoption of domain-specific distributed data
        processing patterns~\parencite{Foidl2024DataDevelopers}.
  \end{itemize}
  
  The field study by~\citeauthor*{Hill2016TrialsStudy} discusses the
  impact of such complexity for practitioners~\parencite{Hill2016TrialsStudy}.
  It was held with 11 participants with at least two years of experience
  with ML.
    Seasoned engineers described that developing MLES
    \Quote{required skills held only by certain `high priests'},
    while debugging them was
    \Quote{something akin to magic and even voodoo}%
  ~\parencite{Hill2016TrialsStudy}.
  Such challenges become even more critical since MLES need
  to be updated and improved continuously~\parencite{Wan2021HowPractices}.
  
  The many surveys in the literature show a general lack of studies regarding
  \emph{metrics} for MLES. Recently, \citeauthor*{Warnett2025AArchitectures}
  were one of the first to publish about the subject, proposing a set of metrics
  to measure how automated is an MLES. Henceforth, this research intends to
  further explore this gap, thus understanding how metrics can help to manage
  the architectural complexity of MLES.

\section{Research Questions}\label{sec:research_questions}

  To achieve the goal of this research, this PhD proposes
  two main research questions. They follow the SMART principle%
  ~\parencite{Verschuren2010DesigningDesign}, i.e., they should be
  \emph{Specific, Measurable, Achievable, Relevant, and Time-Bound}.

  \begin{researchquestion}
    What are the measurable dimensions of complexity in
    the architecture of MLES?
  \end{researchquestion}

  \Cref{rq:1} aims to explore metrics related to complexity available in the
  literature. \emph{Software Metrics} have been thoroughly researched by
  Software Engineering~\parencite{Fenton2014SoftwareEdition}.
  However, MLES also have their data and model dimensions%
  ~\parencite{Sato2019ContinuousLearning},
  which affect the software architecture.
  Finding metrics that measure the data- and model-related complexity 
  \emph{beyond code} is the biggest potential challenge for
  answering~\cref{rq:1}.

  \begin{researchquestion}
    How can complexity metrics be operationalized over
    the architecture of MLES?
  \end{researchquestion}

  \Cref{rq:2} aims to define a process to collect the metrics found
  in \cref{rq:1}.
  Metrics may be related to different abstraction levels of a system:
  from code snippets, to components, to services%
  ~\parencite{Fenton2014SoftwareEdition}.
  Moreover, they can have varying levels of different quality attributes%
  ~\parencite{Latva-Koivisto2001FindingModels,Polancic2017ComplexityReview},
  such as validity, reliability, computability, intuitiveness,
  ease of implementation, and independence of other metrics.
  As a consequence, only some metrics found in \cref{rq:1} may
  be practical or useful to collect. In particular, if two metrics
  provide similar information, it will be preferable to use the best
  according to the quality attributes. 
  Therefore, the challenge for answering~\cref{rq:2} is twofold: creating a
  dataset of representative MLES to collect the metrics, and then choosing which
  subset of metrics to collect.
  
  Measuring a metric may require different levels of access to the system
  (e.g., having the codebase available for processing) or rely on different
  representations of the system (e.g., creating a graph describing its data
  flow). As a consequence, only some metrics found in \cref{rq:1} may be
  practical to collect. In particular, if two metrics provide similar
  information, it may be preferable to use the simpler.

\section{Methodology}\label{sec:research_methodology}

\begin{figure}[ht]
    \centering
    \includegraphics[width=\linewidth]{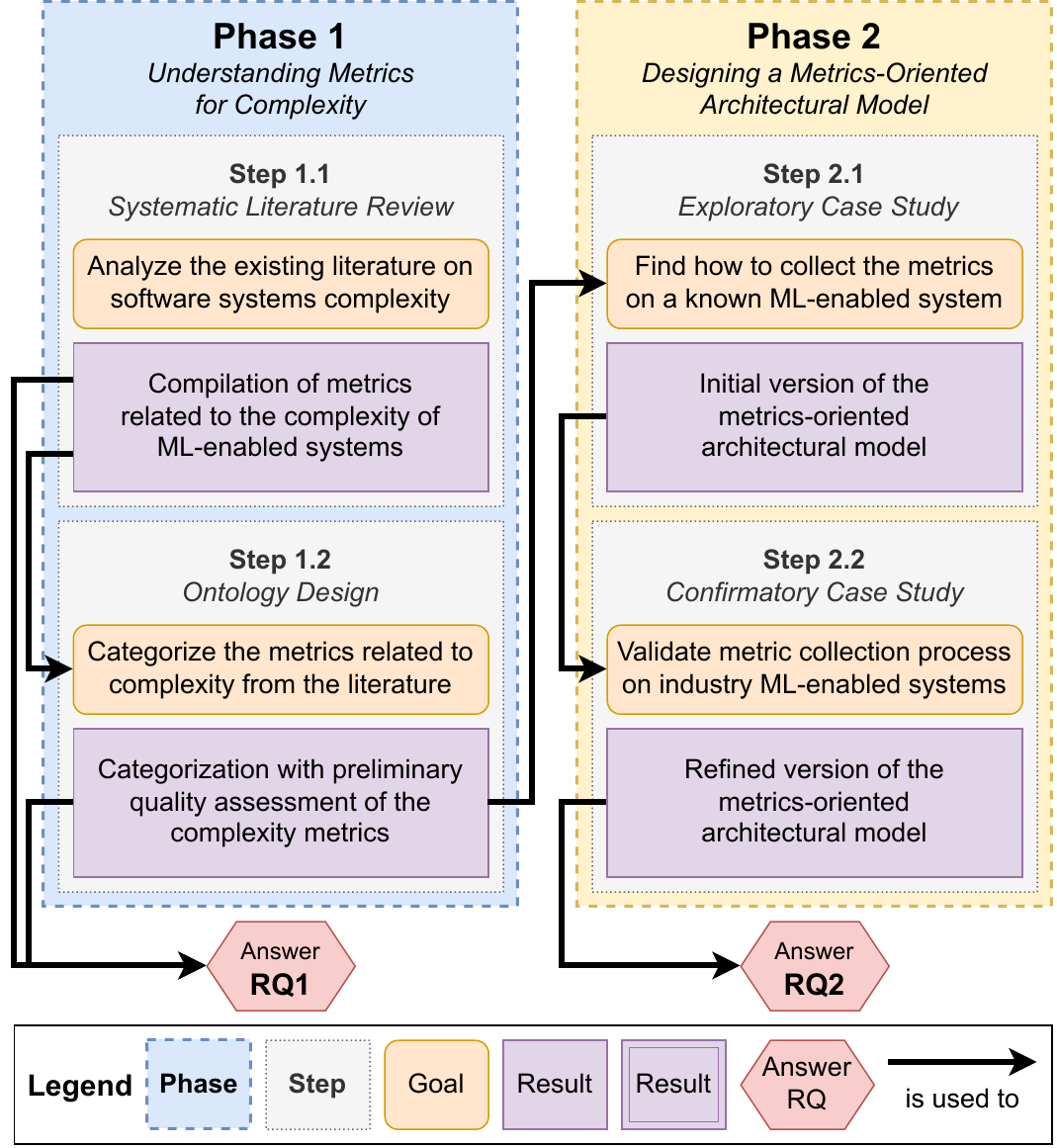}
    \caption[%
      Research Methodology%
    ]{%
      \emph{Research Methodology}.
      The methodology is divided into two phases, each addressing a
      research question from \cref{sec:research_questions}.
      This methodology categorizes as empirical software engineering
      with a mixed-method approach.
    }
    \vspace{-0.25cm}
    \label{fig:research_methodology}
  \end{figure}

  \Cref{fig:research_methodology} summarizes four steps, grouped into two
  phases, of the methodology proposed for this research.

  \subsection{Understanding Metrics for Complexity}
  \label{subsec:methodology_phase_1}
  \MethodologyPhase

  To answer \cref{rq:1}, this research will start studying
  \emph{the complexity of software systems} and \emph{metrics}.
  For that, \cref{phase:1} has been divided into two steps:
    (1) to analyze the existing literature on software systems complexity; and
    \MethodologyStep
    (2) to categorize the metrics related to complexity from the literature.
    \MethodologyStep

  \medskip
  \subsubsection{Systematic Literature Review (SLR)}
  Making an SLR ensures a reproducible way to analyze the literature%
  ~\parencite{Kitchenham2012SystematicEngineering},
  as it is the goal of \cref{step:1.1}. Initially, the SLR will focus on
  finding references about \emph{complexity metrics}, i.e., metrics that
  were created to measure some aspect of complexity. On the one hand,
  this subject has been thoroughly explored by the software engineering
  literature~\parencite{Fenton2014SoftwareEdition}.
  On the other hand, MLES have their data and model dimensions%
  ~\parencite{Alves2024PracticesReview},
  whose complexity reflects into code, but goes beyond it%
  ~\parencite{ElAlaoui2019BigApproaches}.
  Therefore, characterizing the complexity of MLES will
  require looking into more metrics than for traditional software-based
  systems.

  \medskip
  \subsubsection{Ontology Design}
  Making an ontology results in a well-defined categorization of the metrics%
  ~\parencite{NoyOntologyOntology},
  as it is the goal of \cref{step:1.2}. The ontology will focus on describing
  different \emph{types} of metrics~\parencite{Fenton2014SoftwareEdition},
  as well as possible \emph{quality attributes} associated with them%
  ~\parencite{Latva-Koivisto2001FindingModels}. Some properties will be
  related to the \emph{applicability} of the metrics, which will be explored
  only on~\cref{phase:2}. Therefore, the knowledge base will be fully complete
  only by the end of the research.

  An ontology can be created via \emph{ontology design}%
  ~\parencite{NoyOntologyOntology}. An ontology
  is the structure behind a knowledge base, including \emph{classes},
  \emph{properties}, and \emph{restrictions}~\parencite{NoyOntologyOntology}.
  The ontology will focus on describing different \emph{types} of
  metrics~\parencite{Fenton2014SoftwareEdition}, as well as possible
  \emph{quality attributes} associated with them%
  ~\parencite{Latva-Koivisto2001FindingModels}. 

  \subsection{Designing a Metrics-Oriented Architectural Model}
  \label{subsec:methodology_phase_2}
  \MethodologyPhase

  To answer \cref{rq:2}, this research will focus on selecting a subset of the
  metrics from \cref{phase:1} to compose the \emph{metrics-oriented architectural
  model}. For that, \cref{phase:2} has been divided into two steps:
    (1) to find how to collect the metrics on a known MLES; and
    \MethodologyStep
    (2) to validate the metric collection process on industry MLES.
    \MethodologyStep

  \medskip
  \subsubsection{Exploratory Study Case}
  Making an exploratory case study provides a framework to propose an answer
  to~\cref{rq:2}~\parencite{Easterbrook2008SelectingResearch}.
  The exploratory case study will have the following study proposition:
  \emph{``What is a process to operationalize the collection of complexity
  metrics on an MLES?''}, as it is the goal of \cref{step:2.1}.
  To achieve that, it will rely on a \emph{typical case}%
  ~\parencite{Easterbrook2008SelectingResearch}:
  the SPIRA ML-enabled system, explored by this PhD research in previous
  publications~\parencite{Ferreira2022SPIRA:Detection}.

  The familiarity with the codebase should allow the authors to focus
  on operationalizing the metric collection rather than learning how the
  project works. The goal is twofold:
    document \emph{the process of metric collection},
    to achieve the study proposition; and
    determine \emph{the quality attributes of the metrics},
    to refine the ontology made on \cref{step:1.2}.

  By choosing which metrics are worth collecting, and documenting
  the requirements to collect them, the result will be the initial version
  of the \emph{metrics-oriented architectural model}.

  \medskip
  \subsubsection{Confirmatory Study Case}
  Making a confirmatory case study provides a framework to validate the proposed
  answer to~\cref{rq:2}~\parencite{Easterbrook2008SelectingResearch}.
  The confirmatory case study will repeat the same study proposition:
  \emph{``What is a process to operationalize the collection of complexity
  metrics on an MLES?''}, as it is the goal of \cref{step:2.2}.
  To achieve that, it will rely on industry partners of both the University of
  São Paulo (USP) and the Jheronimus Academy of Data Science (JADS).

  Production MLES can have various architectures, with different combinations
  of components. The diversity of codebases should allow the researchers to
  validate the robustness of the metric collection process. Two types of
  data will be revisited:
    \emph{the process of metric collection},
    to achieve the study proposition; and
    \emph{the quality attributes of the metrics},
    to refine the ontology made on \cref{step:1.2}.

  By reviewing which metrics are worth collecting and the requirements
  to collect them, the result will be a refined version of the
  \emph{metrics-oriented architectural model}.

  \section{Preliminary Results}
  \label{sec:ml_enabled_systems}

\begin{figure*}
    \centering
    \includegraphics[width=0.99\textwidth]{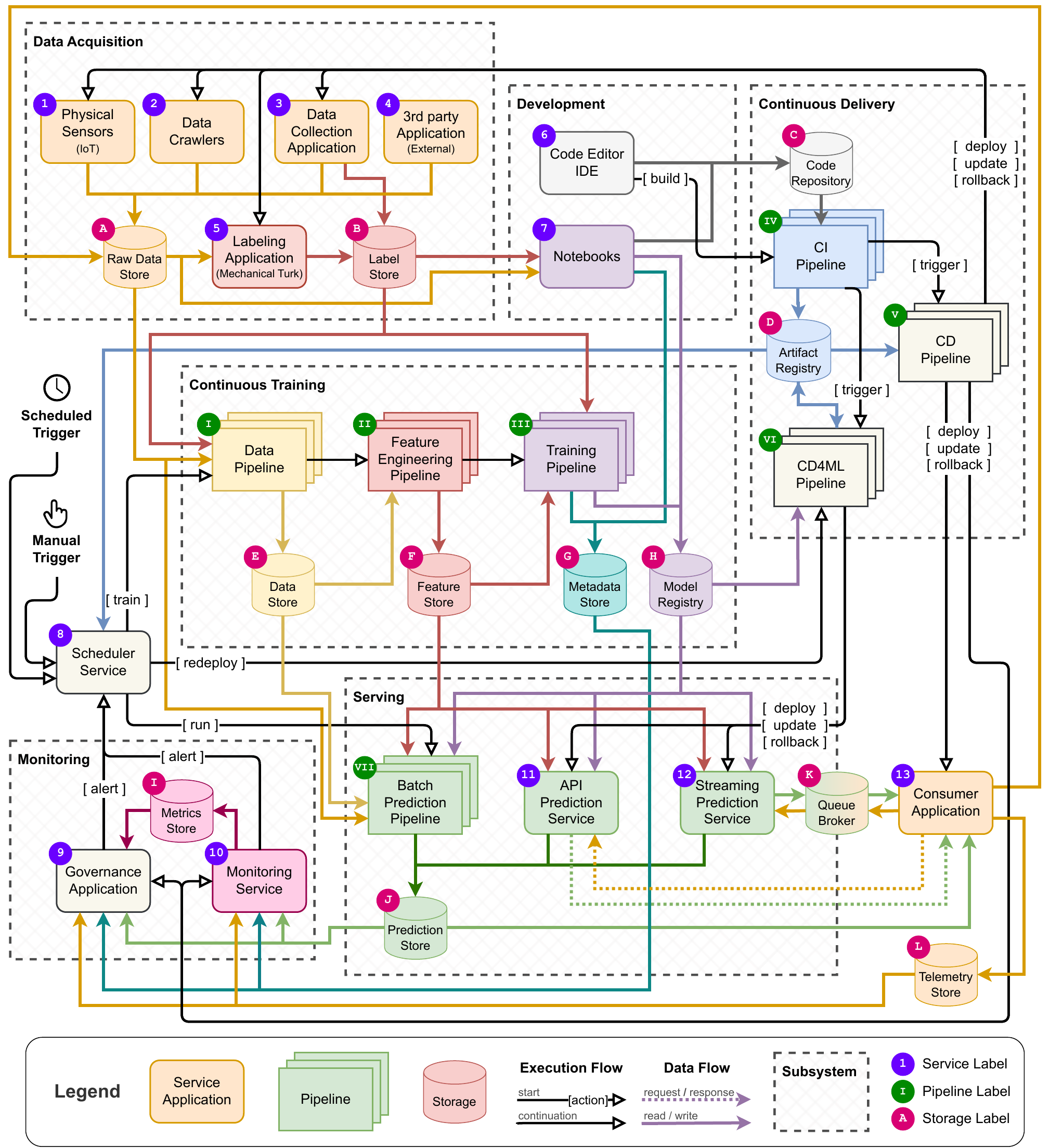}
    \caption[%
      Reference Architecture for ML-enabled Systems
    ]{%
      \emph{Reference Architecture for ML-enabled Systems.}
      There are three types of components in an ML-enabled system.
      Rectangles represent \textbf{applications} or \textbf{services},
        which execute continuously.
      Stacked rectangles represent \textbf{pipelines},
        which execute a task on demand.
      Lastly, cylinders represent \textbf{data storage},
        which may be databases of any type.
      Components are connected by arrows.
      Black arrows with a hollow tip illustrate the \textbf{execution flow}.
        They start and end in a component.
        Labeled arrows represent the trigger that starts a workflow,
        whereas unlabeled arrows represent the continuation of an
        existing workflow.
      Colored arrows with a filled tip illustrate the \textbf{data flow}.
      They appear in two types:
        solid arrows going to and from a data storage represent
        write and read operations, respectively;
        dotted arrows represent a sync or async request-response
        communication between components.
      Components are colored according to the data they produce:
        \mbox{\LegendColoredComponent{orange}{raw data}},
        \mbox{\LegendColoredComponent{red}{ML-specific data}},
        \mbox{\LegendColoredComponent{gray}{source code}},
        \mbox{\LegendColoredComponent{blue}{executable artifacts}},
        \mbox{\LegendColoredComponent{yellow}{ML models}},
        \mbox{\LegendColoredComponent{purple}{ML training metadata}},
        \mbox{\LegendColoredComponent{green}{ML model predictions}}, and
        \mbox{\LegendColoredComponent{pink}{ML model metrics}}.
        Remaining \LegendBWComponent{standalone components} orchestrate
        the execution of others.
      Components are also grouped into \textbf{subsystems}.
      \LegendColoredLabel{violet}{Numbers},
      \LegendColoredLabel{moss}{roman numerals} and
      \LegendColoredLabel{magenta}{letters}
      are used as labels to refer to different components of the reference
      architecture.
      This reference architecture is based on the article by
      \citeauthor*{Kumara2023RequirementsIndustry}
      \parencite{Kumara2023RequirementsIndustry}, which was extended
      with the addition of two new subsystems:
      \textsc{Data Acquisition} and \textsc{Continuous Delivery}.
    }
    \label{fig:reference_architecture}
  \end{figure*}

  \Cref{fig:reference_architecture} presents a reference architecture for MLES.
  It summarizes possible components of an MLES, including \emph{applications and
  services}, \emph{pipelines}, and \emph{data storage}. It groups all components
  into six \emph{subsystems}, which represent the major tasks of an MLES.
  Moreover, it shows how they interact according to the system's
  \emph{execution flow} and \emph{data flow}.
  
  
  The reference architecture is based on
  \citeauthor*{Kumara2023RequirementsIndustry}%
  ~\parencite{Kumara2023RequirementsIndustry}, who proposed its key elements
  using Grounded Theory~\parencite{Wohlin2012ExperimentationEngineering}.
  \Cref{fig:reference_architecture} extends this result
  by introducing two additional subsystems:
    \textsc{data acquisition},
    illustrating data collection components that were not
    addressed in their reference architecture; and
    \textsc{continuous delivery},
    associating some components that were not grouped in
    their reference architecture.

  The preprint by \citeauthor*{Kumara2023RequirementsIndustry}%
  ~\parencite{Kumara2023RequirementsIndustry} has been submitted to
  the \emph{Communications of the ACM}~\parencite{Kumara2024MLOpsIndustry}.
  In 2024, the author contributed to the revision of the paper,
  adding interviews with experts to validate the requirements
  collected by that research~\parencite{Kumara2024MLOpsIndustry}.
  This opportunity was also used to validate the extended reference
  architecture represented in \Cref{fig:reference_architecture}.

  The extended reference architecture will be used to describe systems before
  collecting their metrics. Therefore, it provides the foundation to
  create the \emph{metrics-based architectural model}.

  \section{Expected Results and Threats to Validity}
  \label{sec:methodology_results_threats}

  This section highlights the main expected result (ER) of each phase from
  \Cref{fig:research_methodology}. Then, it presents threats to validity,
  proposing solutions on how to address them.

  \begin{expectedresult}
    State of the art about complexity metrics for MLES.
  \end{expectedresult}

  \Cref{phase:1} is mostly subject to \emph{internal} and \emph{conclusion}
  threats to validity. \Cref{step:1.1} leads to an internal threat because
  the SLR depends on the data provided by knowledge bases. \Cref{step:1.2}
  leads to a conclusion threat because it produces an ontology -- whose
  design is a subjective process by itself -- based on the data gathered
  on the previous step.

  To contain the internal threat, this research intends to produce an SLR
  based on the most established publication databases in Computer Science.
  To contain the conclusion threat, it will follow well-established guidelines
  available in the literature,
  both for the SLR~\parencite{Kitchenham2012SystematicEngineering} and for
  the ontology~\parencite{NoyOntologyOntology}. By documenting the process
  thoroughly, the results from~\cref{phase:1} should be reliable for the
  next phases.

  \begin{expectedresult}
    An academic- and industry-based case study on complexity metrics for
    MLES.
  \end{expectedresult}

  \Cref{phase:2} is mostly subject to \emph{external} and \emph{conclusion}
  threats to validity. Both \cref{step:2.1,step:2.2} are case studies that
  will incrementally result in the \emph{metrics-oriented architectural model}.
  They will be based on the analysis of existing MLES: the SPIRA system for
  \cref{step:2.1}, and production systems from industry partners for
  \cref{step:2.1}. The choice of systems included in the case studies
  leads to the external threat, whereas the selection of metrics that
  leads to the conclusion threat.

  To contain the external threat, this research will define inclusion and
  exclusion criteria to accept an MLES in the case study.
  As it was considered for SPIRA, any production system from an industry
  partner will be considered only if they contain key components of the
  reference architecture explored in~\Cref{fig:reference_architecture}.
  To contain the conclusion threat, \cref{phase:2} was designed to have
  first an exploratory and then a confirmatory case study.

\section{Work Plan}
\label{sec:work_plan}

  This PhD started at University of São Paulo (USP) in 2020, amidst
  the COVID-19 pandemic. The author was employed in a full-time job as
  \emph{Principal ML Engineer}, an experience that inspired this research.
  In 2024, he became a \emph{Scientific Programmer} for the \mbox{MARIT-D}
  project, where results from this research have been applied. Therefore,
  \Cref{tab:work_plan} presents a timeline considering this PhD as a
  part-time activity.
  
  The \emph{metrics-oriented architectural model} will be developed
  iteratively, ensuring its metric collection process is tested in
  multiple systems. By creating a reliable model, the results may
  be able to help industry applications to manage complexity.


  \begin{table}[h!]
    \centering
    \caption{}%
    \vspace{-20pt}
    \includegraphics[width=0.96\linewidth]{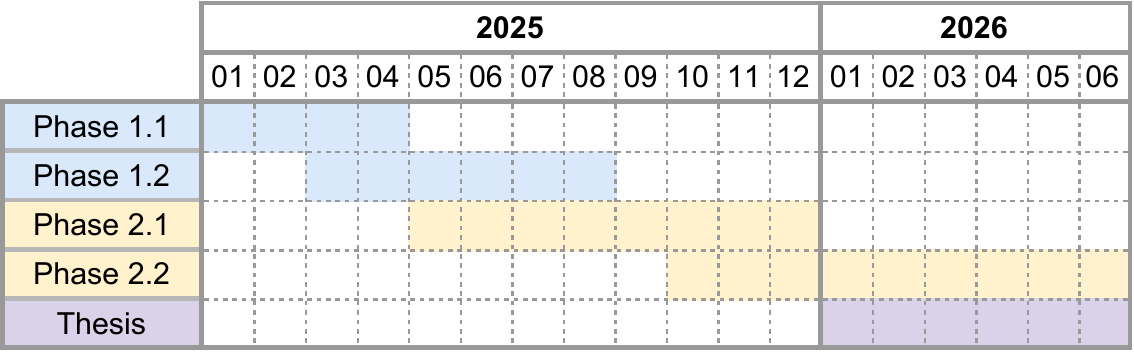}
    \label{tab:work_plan}
  \end{table}


\printbibliography

\end{document}